\documentclass[aps,nofootinbib,notitlepage,showpacs,longbibliography]{revtex4-1}

\usepackage[CJKbookmarks, pdftex, bookmarksnumbered, bookmarksopen, colorlinks, citecolor=blue, linkcolor=blue]{hyperref}

\usepackage{amstext,amsmath,amssymb,amsfonts,bbm}
\usepackage[latin1]{inputenc}
\usepackage{graphicx}
\usepackage{epstopdf}
\usepackage{amsthm}
\usepackage{tocvsec2}
\usepackage{enumerate}
\usepackage{subfigure}
\usepackage{color}

\usepackage{multirow} \usepackage{rotating}

\def\beq{\begin{equation}}
\def\be{\begin{equation}}
\def\ee{\end{equation}}
\def\bes{\begin{eqnarray}}
\def\ees{\end{eqnarray}}

\begin{document}

\title{Curvatures and discrete Gauss-Codazzi equation in (2+1)-dimensional loop quantum gravity}

\author{Seramika Ariwahjoedi$^{1,3}$, Jusak Sali Kosasih$^{3}$, Carlo Rovelli$^{1,2}$, Freddy P. Zen$^{3}$\vspace{1mm}}

\affiliation{$^{1}$Aix Marseille Universit\'e, CNRS, CPT, UMR 7332, 13288 Marseille, France.\\
$^{2}$Universit\'e de Toulon, CNRS, CPT, UMR 7332, 83957 La Garde, France.\\$^{3}$Institut Teknologi Bandung, Bandung 40132, West Java, Indonesia.}

\begin{abstract} 

\noindent
We derive the Gauss-Codazzi equation in the holonomy and plane-angle representations and we use the result to write a Gauss-Codazzi equation for a discrete (2+1)-dimensional manifold, triangulated by isosceles tetrahedra. This allows us to write operators  acting on spin network states in (2+1)-dimensional loop quantum gravity, representing the 3-dimensional intrinsic, 2-dimensional intrinsic, and 2-dimensional extrinsic curvatures.
\end{abstract}

\maketitle

\section{Introduction}

Three distinct notions of curvature are used in general relativity: the intrinsic curvature of the spacetime  manifold, $M$, the intrinsic curvature of the spacial  hypersurface $\Sigma$ embedded in $M$ which is utilised in the canonical framework, and the extrinsic curvature of $\Sigma$  \cite{key-1}.  The three are related by the Gauss-Codazzi equation. On a discrete geometry, the definition of  extrinsic curvature is not entirely clear \cite{key-2}. Even less so in loop quantum gravity (LQG), where the phase-space variables are derived from the first order formalism \cite{key-3}, which generically does not yield a direct interpretation as spacetime geometry. But a definition of these curvatures is important in LQG,  because we expect a discrete (twisted) geometry to emerge from the theory in an appropriate semi-classical limit \cite{key-4}. 

In $n$-dimensional discrete geometry, the manifold is formed by $n$-simplices \cite{key-2,key-5}. The intrinsics curvature sits on the $(n-2)$-simplices, called \textit{hinges}, and can be defined by the \textit{angle of rotation}: a vector parallel-transported around the hinge gets rotated by this angle. The rotation is in the hyperplane dual to the hinge.

In the canonical formulation of general relativity it is convenient to use the ADM formalism \cite{key-6} or a generalisation \cite{key-8}.  The phase space variables are defined on an $n-1$ surface: the ``initial time", or, more generally, ``boundary" surface $\Sigma$. At the core of the ADM formalism is the Gauss-Codazzi equation, relating the intrinsic curvature of the 4D spacetime with the intrinsic and extrinsic curvatures of $\Sigma$. In the context of a discrete geometry, $\Sigma$ is defined as an $(n-1)$-dimensional simplicial manifold formed by $(n-1)$-simplices.   To develop a formalism analogous to the ADM one, we need an equation relating the intrinsic and extrinsic curvatures of $\Sigma$ and $M$, a 'discrete' version of the Gauss-Codazzi equation.

In this paper, we explore a definition of extrinsic curvature for the discrete geometry which allows us to write an equation relating the extrinsic and intrinsics curvatures, which we refer as the Gauss-Codazzi equation for discrete geometry. 

To get some insight into this problem, we first review the standard Gauss-Codazzi equation, both in second and first order formalism. In Section \ref{II} we derive the holonomy (matrix) representation and plane-angle representation of the Gauss-Codazzi equation. We consider the discrete version of the Gauss-Codazzi equation in Section \ref{III}, where we define all the curvatures. In Section \ref{IV}, we move to the loop quantum gravity picture and study the operators corresponding to these curvatures into operators. 

\section{Gauss-Codazzi equation} \label{II}

\subsection{Standard Gauss-Codazzi equation}

The Gauss-Codazzi equation relates the Riemann \textit{intrinsic} curvatures of a manifold and its submanifold with the \textit{extrinsic}
curvature of the submanifold. We will briefly review the continuous (2+1)-dimensional Gauss-Codazzi equation in this section,
but the formula will be valid in $(n+1)$-dimension. 

\subsubsection{Second order formulation}

Consider a 3-dimensional manifold $M$ and a 2-dimensional surface $\Sigma$ embedded in $M$. In the second order formulation
of general relativity \cite{key-7}, the Riemann intrinsic curvature of $M$ is a $\left(\begin{array}{c}
1\\
3
\end{array}\right)$ tensor: 
\[
\,^{3}R=\,^{3}R_{\mu\nu\beta}^{\alpha}\partial_{\alpha}\wedge dx^{\beta}\otimes dx^{\mu}\wedge dx^{\nu},\qquad\alpha,\beta,\mu,\nu=0,1,2
\]
which can be thought as a map that rotates a vector $v\in T_{p}M$ parallel-transported along an infinitesimal square loop defined by unit vectors $\left\{ \partial_{\mu},\partial_{\nu}\right\} $. Next, we define the \textit{projection tensor} as: 
\[
q_{\mu\nu}=g_{\mu\nu}-\left\langle n,n\right\rangle n_{\mu}n_{\nu},
\]
with $n_{\mu}$ is the normal to hypersurface $\Sigma\subset M$. $\left\langle n,n\right\rangle =\pm1$, depends on the signature of
the metric. The projected Riemann curvature of $\,^{3}R$ on $\Sigma$ is defined as: 
\[
\left.\,^{3}R\right|_{\Sigma}=\,^{3}\widetilde{R}_{\mu\nu\beta}^{\alpha}\partial_{\alpha}\wedge dx^{\beta}\otimes dx^{\mu}\wedge dx^{\nu}=q_{\alpha'}^{\alpha}q_{\mu}^{\mu'}q_{\nu}^{\nu'}q_{\beta}^{\beta'}\,^{3}R_{\mu'\nu'\beta'}^{\alpha'}\partial_{\alpha}\wedge dx^{\beta}\otimes dx^{\mu}\wedge dx^{\nu}.
\]
Taking the projected part $\left.\,^{3}R\right|_{\Sigma}$ from the
full part $\,^{3}R$, we have relation as follow: 
\begin{equation}
\,^{3}R=\left.\,^{3}R\right|_{\Sigma}+S,\qquad\left(\textrm{decomposition formula}\right),\label{eq:2.1}
\end{equation}
$S$ is the 'residual' part of $\,^{3}R.$ The projected part $\left.\,^{3}R\right|_{\Sigma}$
can be written as: 
\begin{equation}
\left.\,^{3}R\right|_{\Sigma}=\,^{2}R+\left[K,K\right]\qquad\left(\textrm{Gauss equation}\right),\label{eq:2.2}
\end{equation}
which can be written in terms of components: 
\begin{equation}
\,^{3}\widetilde{R}_{\mu\nu\beta}^{\alpha}=\,^{2}R_{\mu\nu\beta}^{\alpha}+\left(K_{\mu}^{\alpha}K_{\beta\nu}-K_{\nu}^{\alpha}K_{\beta\mu}\right),\quad K_{\alpha\beta}=q_{\alpha}^{\mu}q_{\beta}^{\nu}\nabla_{\mu}n_{\nu}.\label{eq:2.3}
\end{equation}
$\,^{2}R$ and $K$ are, respectively, the 2-dimensional Riemannian
curvature and extrinsic curvature of hypersurface $\Sigma$. The residual
part $S$ can be written as:
\[
S_{\mu\nu\beta}^{\alpha}=n^{\alpha}\left(\,^{2}\nabla_{\nu}K_{\mu\beta}-\,^{2}\nabla_{\mu}K_{\nu\beta}\right)\qquad\left(\textrm{Codazzi equation}\right),
\]
with $\,^{2}\nabla$ is the covariant derivative on the slice $\Sigma.$

Writing (\ref{eq:2.1}) and (\ref{eq:2.2}) together, we obtain: 
\[
\,^{3}R=\underset{\left.\,^{3}R\right|_{\Sigma}}{\underbrace{\,^{2}R+\left[K,K\right]}+S}.
\]

\subsubsection{First order formulation}

The gravitational field is a gauge field and can be written in a form closer to Yang-Mills theory. This way of representing gravity
is known as first order formulation of general relativity \cite{key-3,key-7}. Let manifold $M$ be a spacetime. Let $e$ be a local \textit{trivialization}, a diffeomorphism map between the trivial vector bundle $M\times\mathbb{R}^{3}$ with the tangent bundle over $M$: $TM=\cup_{p}\left\{ p\times T_{p}M\right\}$, and $A$ be the connection on $M\times\mathbb{R}^{3}$. The 3-dimensional intrinsic curvature of the connection is the \textit{curvature 2-form}: 
\[
\,^{3}F=F_{\mu\nu J}^{I}\xi_{I}\wedge\xi^{J}\otimes dx^{\mu}\wedge dx^{\nu},\quad\mu,\nu,I,J=0,1,2.
\]
which comes from the exterior covariant derivative of the connection:
\begin{equation}
F=d_{D}A.\label{eq:2.4}
\end{equation}
$\left\{ \partial_{\mu},dx^{\mu}\right\}$ and $\left\{ \xi_{I},\xi^{I}\right\}$ are local coordinate basis on $M$ and $\mathbb{R}^{3},$ respectively. In terms of components: 
\begin{equation}
F_{\mu\nu J}^{I}=\partial_{\mu}A_{\nu J}^{I}-\partial_{\nu}A_{\mu J}^{I}+A_{\mu K}^{I}A_{\nu J}^{K}-A_{\nu K}^{I}A_{\mu J.}^{K}\label{eq:2.5}
\end{equation}
We use nice coordinates such that the time coordinate $\xi_{0}$ in fibre $\mathbb{R}^{3}$ is mapped by $e$ to the time coordinate $\partial_{0}$ in the basespace $M$, i.e., the map $e$ is fixed into:
\[
e\left(\xi^{0}\right)=e^{0}=\delta_{\mu}^{0}dx^{\mu}=dx^{0}.
\]
For the foliation, we use the \textit{time gauge}, where the normal of the hypersurface $\Sigma$ is taken to be the time direction:
\[
n=\partial_{0},\qquad e^{-1}\left(n\right)=\xi_{0}.
\]
Then the $(2+1)$ split is simply carried by spliting the indices
as $I=0,a,$ and $\mu=0,i,$ with $0$ as the temporal part: 
\begin{eqnarray}
\,^{3}F & = & \left(-F_{a0i0}+F_{00ia}+F_{ai00}-F_{0i0a}\right)\xi^{a}\wedge\xi^{0}\otimes dx^{i}\wedge dx^{0}+\left(F_{aij0}-F_{0ija}\right)\xi^{a}\wedge\xi^{0}\otimes dx^{i}\wedge dx^{j}+\label{eq:2.5a}\\
 &  & +\left(F_{ai0b}-F_{a0ib}\right)\xi^{a}\wedge\xi^{b}\otimes dx^{i}\wedge dx^{0}+F_{aijb}\xi^{a}\wedge\xi^{b}\otimes dx^{i}\wedge dx^{j}\nonumber 
\end{eqnarray}
The projected part of the curvature 2-form on $\Sigma$ is: 
\[
\left.\,^{3}F\right|_{\Sigma}=F_{aijb}\xi^{a}\wedge\xi^{b}\otimes dx^{i}\wedge dx^{j},
\]
and the residual part is: 
\begin{eqnarray*}
S & = & \left(-F_{a0i0}+F_{00ia}+F_{ai00}-F_{0i0a}\right)\xi^{a}\wedge\xi^{0}\otimes dx^{i}\wedge dx^{0}+\left(F_{aij0}-F_{0ija}\right)\xi^{a}\wedge\xi^{0}\otimes dx^{i}\wedge dx^{j}\\
 &  & +\left(F_{ai0b}-F_{a0ib}\right)\xi^{a}\wedge\xi^{b}\otimes dx^{i}\wedge dx^{0}.
\end{eqnarray*}
Therefore, the \textit{decomposition formula} is clearly: 
\begin{equation}
\,^{3}F=\left.\,^{3}F\right|_{\Sigma}+S.\label{eq:2.6}
\end{equation}
Next, we take only the projected part: 
\[
\left.\,^{3}F\right|_{\Sigma}=F_{aijb}\xi^{a}\wedge\xi^{b}\otimes dx^{i}\wedge dx^{j},
\]
which can be written in terms of components as follow: 
\begin{equation}
F_{aijb}=\,^{2}F_{aijb}+A_{i0}^{a}A_{jb}^{0}-A_{j0}^{a}A_{ib}^{0}\qquad\textrm{(Gauss equation).}\label{eq:2.7}
\end{equation}
The closed part of $\left.\,^{3}F\right|_{\Sigma}$ is clearly the
3D intrinsic curvature of connection in $\Sigma$ and the rest is
the extrinsic curvature part: 
\[
\,^{2}F=\,^{2}F_{aijb}\xi^{a}\wedge\xi^{b}\otimes dx^{i}\wedge dx^{j}=\left(\partial_{i}A_{ajb}-\partial_{j}A_{aib}+A_{aic}A_{jb}^{c}-A_{ajc}A_{ib.}^{c}\right)\xi^{a}\wedge\xi^{b}\otimes dx^{i}\wedge dx^{j},
\]
\[
K=K_{i}^{a}\xi_{\alpha}\otimes dx^{i}=A_{i0}^{a}\xi_{\alpha}\wedge\xi^{0}\otimes dx^{i}.
\]
The residual part $S,$ in this special coordinates and special gauge
fixing satisfies:
\[
F_{0ija}=\,^{2}D_{j}K_{ia}-\,^{2}D_{i}K_{ja}\qquad\textrm{(Codazzi equation),}
\]
and the other temporal components are zero. It must be kept in mind that this is the Gauss-Codazzi equation for a special local coordinate
and special choice of gauge fixing, for a general case, they are not this simple.

To conclude, we have the decomposition and Gauss-Codazzi equation for a fibre bundle of gravity: 
\[
\,^{3}F=\underset{\left.\,^{3}F\right|_{\Sigma}}{\underbrace{\,^{2}F+\left[\,^{2}K,\,^{2}K\right]}+S.}
\]

\subsection{Gauss-Codazzi equation in holonomy representation}

\subsubsection{Holonomy around a loop}

In this section, we will write the Gauss-Codazzi equation in terms
of holonomy. The holonomy is defined by the parallel transport of
any section of a bundle, say, $s_{0}$, so that it satisfies the equation
as follow: 
\begin{eqnarray}
D_{v}\left(s_{0}\right) & =\partial s_{0}+A\left(s_{0}\right)= & 0.\label{eq:2.8}
\end{eqnarray}
Solving (\ref{eq:2.8}) using recursive method \cite{key-3,key-7, key-8}, we obtain
the solution: 
\[
s\left(t\right)=U\left(\gamma\left(t\right),D\right)s_{0},
\]
$U\left(\gamma\left(t\right),D\right)$ is\textit{ holonomy} of connection
$D$ along path $\gamma\left(t\right)$: 
\begin{equation}
U\left(\gamma\left(t\right),D\right)=U_{\gamma}\equiv{P}\exp\int A,\label{eq:2.30}
\end{equation}
with ${P}$ is the \textit{path-ordered operator} (See \cite{key-3,key-8} for the details of the derivation).

Consider a square loop $\gamma$ embedded in $\Sigma$
which encloses a 2-dimensional area. The holonomy around the square
loop can be written as a product of four holonomies, since holonomy
is piecewise-linear: 
\[
U_{\gamma}=U_{\gamma_{4}}U_{\gamma_{3}}U_{\gamma_{2}}U_{\gamma_{1},}
\]
Taylor expanding the holonomy in (\ref{eq:2.30}) up to the second
order \cite{key-3,key-7, key-9}, we obtain: 
\[
U_{\gamma}=1+\frac{a^{\mu\nu}}{2}F_{\mu\nu}+\mathcal{O}^{4},
\]
with $a^{\mu\nu}$ is an infinitesimal square area inside loop $\gamma.$
The formula: 
\begin{equation}
\frac{a^{\mu\nu}}{2}F_{\mu\nu}=U_{\gamma}-1-\mathcal{O}^{4}\label{eq:2.31}
\end{equation}
will be used to write the Gauss-Codazzi equation in terms of holonomies.

\subsubsection{First order formulation in holonomy representation}

Contracting (\ref{eq:2.5a}) by an infinitesimal area $a^{\mu\nu}$,
we obtain: 
\begin{eqnarray*}
\frac{a^{\mu\nu}}{2}\,^{3}F_{I\mu\nu J} & =\,^{3}U_{IJ}-\delta_{IJ}-\mathcal{O}^{4}= & \frac{a^{i0}}{2}\left(-F_{a0i0}+F_{00ia}+F_{ai00}-F_{0i0a}\right)\xi^{a}\wedge\xi^{0}+\frac{a^{ij}}{2}\left(F_{aij0}-F_{0ija}\right)\xi^{a}\wedge\xi^{0}+\\
 &  & +\frac{a^{i0}}{2}\left(F_{ai0b}-F_{a0ib}\right)\xi^{a}\wedge\xi^{b}+\frac{a^{ij}}{2}F_{aijb}\xi^{a}\wedge\xi^{b},
\end{eqnarray*}
where we have used the result in (\ref{eq:2.31}), relating the holonomy
with the curvature of the connection. Taking the projected part (\ref{eq:2.7})
contracted with $a^{ij},$ and using the relation between holonomy
with the curvature 2-form, we can write the projected part in terms
of 2-dimensional holonomy and the contracted extrinsic curvature as
follow: 
\begin{eqnarray*}
\frac{a^{ij}}{2}F_{aijb} & = & \frac{a^{ij}}{2}\,^{2}F_{aijb}+\frac{a^{ij}}{2}\left(K_{ai}K_{jb}-K_{aj}K_{ib}\right)=\,^{2}U_{ab}-\delta_{ab}-\mathcal{O}^{4}+\frac{a^{ij}}{2}\left(K_{ai}K_{jb}-K_{aj}K_{ib}\right).
\end{eqnarray*}
Finally, collecting all these result together, we obtain: 
\begin{eqnarray*}
\underset{\left(\,^{3}U_{IJ}-\delta_{IJ}-\mathcal{O}^{4}\right)\xi^{I}\wedge\xi^{J}}{\underbrace{\frac{a^{\mu\nu}}{2}\,^{3}F_{I\mu\nu J}\xi^{I}\wedge\xi^{J}}} & = & \underset{\left(\,^{2}U_{ab}-\delta_{ab}-\mathcal{O}^{4}\right)\xi^{a}\wedge\xi^{b}}{\underbrace{\frac{a^{ij}}{2}\,^{2}F_{aijb}\xi^{a}\wedge\xi^{b}}}+\underset{\left[K,K\right]_{ab}\xi^{a}\wedge\xi^{b}}{\underbrace{\frac{a^{ij}}{2}\left(K_{ai}K_{jb}-K_{aj}K_{ib}\right)\xi^{a}\wedge\xi^{b}}}\\
 &  & \left.\begin{array}{c}
+\frac{a^{i0}}{2}\left(\left(-F_{a0i0}+F_{00ia}+F_{ai00}-F_{0i0a}\right)\xi^{a}\wedge\xi^{0}+\left(F_{ai0b}-F_{a0ib}\right)\xi^{a}\wedge\xi^{b}\right)\\
+\frac{a^{ij}}{2}\left(F_{aij0}-F_{0ija}\right)\xi^{a}\wedge\xi^{0}
\end{array}\right\} S,
\end{eqnarray*}
or simply: 
\begin{equation}
\,^{3}U=\,^{2}U+\underset{T}{\underbrace{\left[K,K\right]+S+\left(\,^{3}I-\,^{2}I\right)}},\label{eq:2.9}
\end{equation}
where we have written the Gauss-Codazzi and the decomposition formula together,
noting the residual terms as $S$. Remember that $U$ is the holonomy
of the connection, which describes the curvature of the connection
of fibre $F$, \textit{not} the curvature of the basespace $M$.

\subsubsection{Second order formulation in holonomy representation}

In the same manner as above, for a tangent bundle $TM$, we obtain the relation 
\begin{eqnarray*}
\underset{\left(\,^{3}H_{\alpha\beta}-\delta_{\alpha\beta}-\mathcal{O}^{4}\right)dx^{\alpha}\wedge dx^{\beta}}{\underbrace{\frac{a^{\mu\nu}}{2}\,^{3}R_{\alpha\mu\nu\beta}dx^{\alpha}\wedge dx^{\beta}}} & = & +\underset{\left(\,^{2}H_{kl}-\delta_{kl}-\mathcal{O}^{4}\right)dx^{k}\wedge dx^{l}}{\underbrace{\frac{a^{ij}}{2}\,^{2}R_{kijl}dx^{k}\wedge dx^{l}}}+\underset{\left[K,K\right]_{kl}dx^{k}\wedge dx^{l}}{\underbrace{\frac{a^{ij}}{2}\left(K_{ki}K_{jl}-K_{kj}K_{il}\right)dx^{k}\wedge dx^{l}}}\\
 &  & \left.\begin{array}{c}
+\left(\frac{a^{i0}}{2}\left(-R_{j0i0}+R_{00ij}+R_{ji00}-R_{0i0j}\right)+\frac{a^{ij}}{2}\left(R_{jik0}-R_{0ikj}\right)\right)dx^{j}\wedge dx^{0}\\
+\frac{a^{i0}}{2}\left(R_{ji0k}-R_{j0ik}\right)dx^{j}\wedge dx^{k}
\end{array}\right\} S,
\end{eqnarray*}
or simply: 
\begin{equation}
\,^{3}H=\,^{2}H+\underset{T}{\underbrace{\left[K,K\right]+S+\left(\,^{3}I-\,^{2}I\right)}},\label{eq:2.10}
\end{equation}
with $H$ is the holonomy around any loop $\gamma$ embedded in $\Sigma\subseteq M$, coming from the Riemann tensor $R$.

\subsection{Gauss-Codazzi equation in plane-angle representation}

Rotations can be represented in two ways, using the \textit{holonomy (matrix) representation}, or using the \textit{plane-angle representation}. For a continuous theory, the holonomy representation provides a simpler way to do calculation concerning rotation. In a discrete theory,
where we would like to get rid of coordinates, the plane-angle representation is a natural way to represent rotations and curvatures. The plane-angle representation is only a representation of the rotation group using its Lie algebra for the plane of rotation and one (real) parameter group times the norm of the algebra for the angle of rotation. There exist a bijective map sending the holonomy to the plane-angle representation, known as the \textit{exponential map} \cite{key-10}.

In this section, we rewrite the Gauss-Codazzi equation using the plane-angle representation. We only do the calculation for the second order formalism, the first order formalism version can be obtained in a similar way. Firstly, let us define the variables; we have two equations: the decomposition formula and the Gauss-Codazzi equation written compactly in (\ref{eq:2.10}). Since we are working in (2+1)-dimension, it is natural to use matrix group $SO(3)$ to represents the 3-dimensional intrinsic curvature, and the subgroup $SO(2)$ to represents the 2-dimensional intrinsic curvature. But for simplicity of the calculation, we use the unitary group $SU(2)$ --the double cover of $SO(3)$-- instead of $SO(3)$. Having information of $\,^{3}H\in SU(2)$ is equivalent with having information of the plane \textit{and} angle of the 3-dimensional rotation, which are denoted, respectively, by $\left\{ J\in\mathfrak{su(2)},\delta\theta\in\mathbb{R}\right\} $. The same way goes with $\,^{2}H\in SO(2)$, it contains same amount of information with the plane-angle of the 2-dimensional rotation $\left\{ j\in\mathfrak{so(2)},\delta\phi\in\mathbb{R}\right\}$.
Using the exponential map, we can write: 
\begin{equation}
^{2}H=\exp\left(j\delta\phi\right)=I_{2\times2}\cos\delta\phi+j\sin\delta\phi,\label{eq:2-10a}
\end{equation}
with 
\begin{equation}
j=\left[\begin{array}{cc}
0 & 1\\
-1 & 0
\end{array}\right],\label{eq:2.11a}
\end{equation}
is the generator of $SO(2).$ Doing the same way to the element of
$SU(2)$, we obtain: 
\begin{equation}
^{3}H\sim\exp\left(J\delta\theta\right)=I_{2\times2}\cos\frac{\delta\theta}{2}+J\sin\frac{\delta\theta}{2},\label{eq:2-11}
\end{equation}
with $J\in\mathfrak{su(2)}$ is an element of the Lie algebra of SU(2),
satisfying: 
\[
J=J^{a}\bar{\sigma}_{a},\quad J^{a}J_{a}=1.
\]
$\bar{\sigma}_{a}$ are the basis of $\mathfrak{su(2),}$ namely,
the generator of $SU(2)$: 
\[
\bar{\sigma}_{a}=i\sigma_{a},
\]
satisfying the algebra structure relation as follow: 
\begin{equation}
\left[\bar{\sigma}_{a},\bar{\sigma}_{b}\right]=-2\varepsilon_{ab}^{\:\, c}\bar{\sigma}_{c},\label{eq:3}
\end{equation}
with $\sigma_{a}$ are the Pauli matrices: 
\[
\sigma_{x}=\left[\begin{array}{cc}
0 & 1\\
1 & 0
\end{array}\right],\quad\sigma_{y}=\left[\begin{array}{cc}
0 & -i\\
i & 0
\end{array}\right],\quad\sigma_{z}=\left[\begin{array}{cc}
1 & 0\\
0 & -1
\end{array}\right]
\]
The $\frac{1}{2}$ factor in (\ref{eq:2-11}) comes out from the normalization
occuring when we write $SO(3)$ using $SU(2)$ representation, i.e.,
to reduce the factor 2 in relation (\ref{eq:3})

Taking the trace of (\ref{eq:2.10}) gives the relation between 3-dimensional
and 2-dimensional rotation angle: 
\begin{equation}
\underset{2\cos\frac{\delta\theta}{2}}{\underbrace{\textrm{tr}\,^{3}H}}=\underset{2\cos\delta\phi}{\underbrace{\textrm{tr}\,^{2}H}}+\textrm{tr}T,\label{eq:2.12}
\end{equation}
using the fact that the elements of the algebra are skew-symmetric.
The other relation we need to have the full information contained 
by (\ref{eq:2.10}) is the planes of rotation. Since the embedded
surface is 2-dimensional, the plane of rotation for $\delta\phi$
is trivial, which is (\ref{eq:2.11a}). The plane of rotation $J$
for $\delta\theta$ can be obtained by solving complex $2\times2$
matrix linear equation (\ref{eq:2-11}): 
\begin{equation}
J=\frac{^{3}H-I_{2\times2}\cos\frac{\delta\theta}{2}}{\sin\frac{\delta\theta}{2}}.\label{eq:2.13}
\end{equation}
In conclusion, the contracted Gauss-Codazzi equation concerning
the relation of curvatures of a manifold and its submanifold in (2+1)
dimension can be written using holonomy representation (\ref{eq:2.10})
or using the plane-angle representation, i.e., the relation between
rotation angles by (\ref{eq:2.12}) and the condition for the planes
of rotation, by (\ref{eq:2.11a}) and (\ref{eq:2.13}).

\section{Discrete (2+1) geometry} \label{III}

\subsection{Geometrical setting}

Using the Gauss-Codazzi equation in terms of holonomy and plane-angle
representation obtained in the previous section, we can write the Gauss-Codazzi 
equation for discrete (2+1) geometry; this is the objective of this section.
Let a portion of curved 3-dimensional manifold $M$ be discretized
by flat tetrahedra, we call this discretized manifold $M_{\triangle}$.
See FIG. 1.

\begin{figure}
\centerline{\includegraphics[height=4cm]{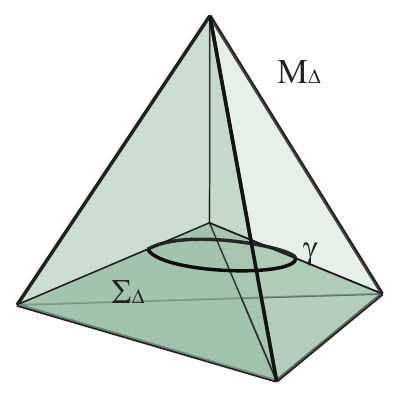}} \caption{Suppose we have a 3-dimensional curved manifold $M_{\triangle}$ discretized
by four tetrahedron in the figure above (in flat case it is known
as \textit{1-4 Pachner move}). Then we take an embedded slice $\Sigma_{\triangle}$
as the surface of one tetrahedron (the dark blue surface discretized
by three triangle). Embedded on $\Sigma_{\triangle},$ we take a loop
$\gamma$ circling a point of a tetrahedron. Attached to $\gamma$,
are the $SU(2)$ and $SO(2)$ holonomy, which are related to the 3D
and 2D intrinsic curvature, respectively.}
\end{figure}

\subsubsection{Angle relation}

In this section we derive the Gauss-Codazzi equation purely from the angles between simplices, i.e, the relation between geometrical objects, without reference to coordinates (this is, indeed, the reason Tullio Regge developed his \textit{Regge calculus} and discrete general relativity, in his famous paper titled 'General relativity without coordinates' \cite{key-5}).  There are two kinds of relevant 
angles in a flat tetrahedron: the \textit{'angle at a vertex'} between two segments  of the tetrahedron meeting at a vertex, 
which we denote $\phi$;  and the \textit{'dihedral angle at a segment'}, namely the angle between two triangles, which we denote $\theta$. 
These two types of angles are related by the 'angle formula of the
tetrahedron': 
\begin{equation}
\cos\theta_{abac}=\frac{\cos\phi_{bc}-\cos\phi_{ab}\cos\phi_{ac}}{\sin\phi_{ab}\sin\phi_{ac}}.\label{eq:3-1}
\end{equation}
See FIG. 2. 
\begin{figure}
\centerline{\includegraphics[height=3.5cm]{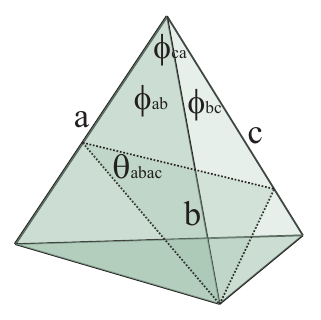}} \caption{Given angles $\phi_{ab}$, $\phi_{bc}$, $\phi_{ca}$ at point $p$
of a tetrahedron, we could obtain the dihedral angle $\theta_{abac}$.
In fact, $\theta_{abac}$ is only $\phi_{bc}$ projected on the plane
normal to segment $a$.}
\end{figure}

\subsubsection{Isosceles tetrahedron}

For simplicity of the derivation in the next section, we consider
\textit{isosceles} tetrahedron. An isosceles tetrahedron is build
by three isosceles triangles and one equilateral triangle. In this
case, all angles around one point $p$ of the tetrahedron are equal, 
say $\phi_{i}=\phi$, and the dihedral angle relation (\ref{eq:3-1})
becomes: 
\begin{eqnarray*}
\cos\theta & = & \frac{\cos\phi}{\left(1+\cos\phi\right)}.
\end{eqnarray*}
This implies that the dihedral angles between two isosceles triangles 
are equal as well, say $\theta_{i}=\theta$. Another geometrical property
of a tetrahedron which is useful for the derivation in this section
is the volume. The volume of unit isosceles parallelepiped is: 
\begin{equation}
\textrm{vol}=\left(\vec{n}_{1}\times\vec{n}_{2}\right)\cdot\vec{n}_{3}=\frac{1}{3}\left(1-\cos\phi\right)\sqrt{1+2\cos\phi}=\frac{1}{3}\left(\sin^{2}\frac{\phi}{2}\right)\sqrt{4\cos^{2}\frac{\phi}{2}-1}.\label{eq:vol}
\end{equation}

\subsection{Curvatures in discrete geometry}

In $n$-dimensional discrete geometry, the notion of curvature is
represented using the plane-angle representation. The intrinsic curvature
is represented by the angle of rotation, which in this case is called
as \textit{deficit angle}, and this deficit angle is located on the
\textit{hinge}, the $(n-2)$ form duals to the plane of rotation.
The geometrical interpretation of deficit angle is illustrated by
the FIG. 3.

\begin{figure}
\centerline{\includegraphics[height=3cm]{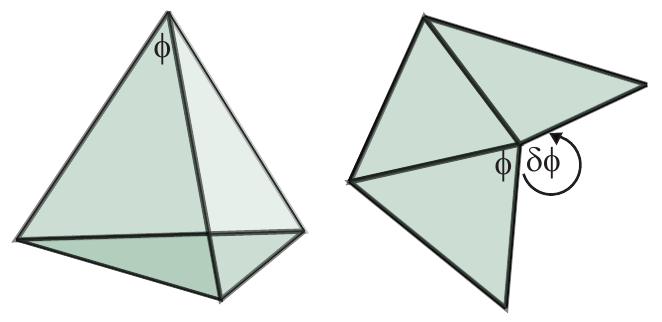}} \caption{Suppose $\Sigma_{\Delta}$ is a 2D surface discretized by three triangles.
Point $p$ is the hinge where the deficit angle $\delta\phi$ is located.
$\delta\phi$ is the 2D intrinsic curvature of $\Sigma_{\Delta}$
at point $p$.}
\end{figure}

To obtain the 2-dimensional and 3-dimensional intrinsic curvatures
in discrete geometry picture, firstly we need to define a discretized
surface embedded in $M_{\Delta},$ say $\Sigma_{\Delta}.$ The simplest
way is to take the surface of the tetrahedron as a slice (see FIG.
1). Having $\Sigma_{\Delta}$ embedded on $M_{\Delta}$,  we consider 
a loop $\gamma$ on $\Sigma_{\Delta}.$  We consider the holonomy 
related to the 2D and 3D curvatures defined by this loop. The
3D curvature is defined by the 3D holonomy $^{3}H\in SO(3)\sim SU(2)$
around loop $\gamma$, which describes the intrinsic curvature of
$M_{\Delta},$ while the 2D curvature is the 2D holonomy $^{2}H\in SO(2),$
describing the intrinsic curvature of slice $\Sigma_{\Delta}$, still
along the same loop. Taking trace of the holonomies, we obtain
the angles of rotation, i.e., \textit{the amount of a components of
a vector parallel to the plane of rotation get rotated when parallel
transported along loop $\gamma$}. In discrete geometry, this rotation
angle is the same as the deficit angle on the hinges.

\subsubsection{2D intrinsic curvature $(SO(2))$}

Using FIG. 4 as the simplest case, we take loop $\gamma$ circling
point $p$ on the surface $\Sigma$, therefore the 2D curvature is
the deficit angle on $p$:
\begin{equation}
^{2}R=\delta\phi=2\pi-\sum_{i}\phi_{i},\label{eq:(aa)}
\end{equation}
or in terms of the cosine function:
\begin{equation}
\cos\delta\phi=\prod_{i=1}^{3}\cos\phi_{i}-\sum_{i=1,i\neq j,k,j<k}^{3}\cos\phi_{i}\sin\phi_{j}\sin\phi_{k}.\label{eq:bb}
\end{equation}
For the isosceles tetrahedron case, (\ref{eq:(aa)})-(\ref{eq:bb})
become: 
\[
^{2}R=\delta\phi=2\pi-3\phi,
\]
\[
\cos\delta\phi=\cos^{3}\phi-3\cos\phi\sin^{2}\phi=\cos\phi\left(4\cos^{2}\phi-3\right).
\]
We could obtain the holonomy by the exponential map (\ref{eq:2-10a}),
using the $SO(2)$ algebra (\ref{eq:2.11a}).

\subsubsection{3D intrinsic curvature $(SU(2))$}

We could write 3D curvature by the holonomy of SU(2) using (\ref{eq:2-11}),
but the 3-dimensional holonomy along loop $\gamma$ circling point
$p$ is a \textit{product} of three holonomies on the three hinges,
because the loop crosses three hinges. See FIG. 4.

\begin{figure}
\centerline{\includegraphics[height=4cm]{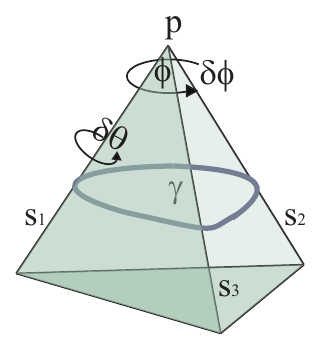}} \caption{A portion of a discretized curved manifold $M_{\Delta}$ with an embedded
discretized hypersurface $\Sigma_{\Delta}$. It is clear that the
loop $\gamma$ circles point $p$, where the 2D deficit angle $\delta\phi$
is located. $\delta\phi$ represents 2D intrinsic curvature since
it is located on a point. The loop also crosses 3 segments which are
$s_{1}$, $s_{2}$, $s_{3}.$ Each segments contains its own 3D deficit
angle $\delta\theta,$ which represent 3D intrinsic curvature since
it is located on a line. Using the exponential map, we can obtain
holonomies on each line corresponding to the deficit angles. From
these holonomies, we can obtain the 3D total holonomy along loop $\gamma$,
which is the product of holonomies on the 3 segments.}
\end{figure}

\begin{equation}
^{3}H=\prod_{i=1}^{3}{}^{3}H_{i}\label{eq:hol}
\end{equation}
Using (\ref{eq:2-11}), we obtain:
\begin{eqnarray*}
^{3}H & = & \prod_{i=1}^{3}\left(I\cos\frac{\delta\theta_{i}}{2}+J_{i}\sin\frac{\delta\theta_{i}}{2}\right)\\
 & = & I\prod_{i=1}^{3}\cos\frac{\delta\theta_{i}}{2}+\sum_{i=1,i\neq j,k,j<k}^{3}J_{i}\sin\frac{\delta\theta_{i}}{2}\cos\frac{\delta\theta_{j}}{2}\cos\frac{\delta\theta_{k}}{2}+\sum_{i=1,i<j,i,j\neq k}^{3}J_{i}\sin\frac{\delta\theta_{i}}{2}J_{j}\sin\frac{\delta\theta_{j}}{2}\cos\frac{\delta\theta_{k}}{2}+\prod_{i=1}^{3}J_{i}\sin\frac{\delta\theta_{i}}{2},
\end{eqnarray*}
 where each hinge (segment) $i$ is dual to plane $J_{i}=\vec{J}_{i}\cdot\hat{\bar{\sigma}}$
and is attached by a deficit angle $\delta\theta_{i}$. Taking trace
of the equation above gives:
\begin{align*}
\underset{2\cos\frac{\delta\Theta}{2}}{\underbrace{\textrm{tr}{}^{3}H}} & =\underset{2}{\underbrace{\textrm{tr}I}}\prod_{i=1}^{3}\cos\frac{\delta\theta_{i}}{2}+\sum_{i=1,i\neq j,k,j<k}^{3}\underset{0}{\underbrace{\textrm{tr}J_{i}}}\sin\frac{\delta\theta_{i}}{2}\cos\frac{\delta\theta_{j}}{2}\cos\frac{\delta\theta_{k}}{2}+\sum_{i=1,i<j,i,j\neq k}^{3}\underset{-2\left(\vec{J}_{i}\cdot\vec{J}_{j}\right)}{\underbrace{\textrm{tr}\left(J_{i}J_{j}\right)}}\sin\frac{\delta\theta_{i}}{2}\sin\frac{\delta\theta_{j}}{2}\cos\frac{\delta\theta_{k}}{2}\\
 & +\underset{+2\left(\vec{J}_{1}\times\vec{J}_{2}\right)\cdot\vec{J}_{3}}{\underbrace{\textrm{tr}\left(\prod_{i=1}^{3}J_{i}\right)}}\prod_{i=1}^{3}\sin\frac{\delta\theta_{i}}{2},
\end{align*}
or simply:
\[
\cos\frac{\delta\Theta}{2}=\prod_{i=1}^{3}\cos\frac{\delta\theta_{i}}{2}-\sum_{i=1,i<j,i,j\neq k}^{3}\left(\vec{J}_{i}\cdot\vec{J}_{j}\right)\sin\frac{\delta\theta_{i}}{2}\sin\frac{\delta\theta_{j}}{2}\cos\frac{\delta\theta_{k}}{2}+\left(\vec{J}_{1}\times\vec{J}_{2}\right)\cdot\vec{J}_{3}\sin\frac{\delta\theta_{i}}{2},
\]
But $\left(\vec{J}_{i}\cdot\vec{J}_{j}\right)=\cos\phi_{i}$, and
$\left(\vec{J}_{1}\times\vec{J}_{2}\right)\cdot\vec{J}_{3}$ is just
the volume of the parallelepiped defined by $\left\{ \vec{J}_{1},\vec{J}_{2},\vec{J}_{3}\right\} $.
(Remember that $\vec{J}=\star J$, or 1-form $\vec{J}$ is the Hodge
dual of the 2-form $J$ in 3-dimensional vector space. The direction
of $\vec{J}$ is normal to the plane discribed by $J$). Therefore,
the 3D intrinsic curvature is:%
\footnote{(\ref{eq:cc}) has a similar form with (\ref{eq:bb}), except the
$\cos\phi_{i}$ and the term which contains the volume form.%
}
\begin{equation}
\cos\frac{\delta\Theta}{2}=\prod_{i=1}^{3}\cos\frac{\delta\theta_{i}}{2}-\sum_{i=1,i<j,i,j\neq k}^{3}\cos\phi_{i}\sin\frac{\delta\theta_{i}}{2}\sin\frac{\delta\theta_{j}}{2}\cos\frac{\delta\theta_{k}}{2}+\left(\textrm{vol}\right)\prod_{i=1}^{3}\sin\frac{\delta\theta_{i}}{2},\label{eq:cc}
\end{equation}
or
\[
^{3}R=\delta\Theta=2\arccos\left(\prod_{i=1}^{3}\cos\frac{\delta\theta_{i}}{2}-\sum_{i=1,i<j,i,j\neq k}^{3}\cos\phi_{i}\sin\frac{\delta\theta_{i}}{2}\sin\frac{\delta\theta_{j}}{2}\cos\frac{\delta\theta_{k}}{2}+\left(\textrm{vol}\right)\prod_{i=1}^{3}\sin\frac{\delta\theta_{i}}{2}\right),
\]
with $\textrm{vol}$ is the volume form described in (\ref{eq:vol}).
For isosceles tetrahedron case, we have $\phi_{i}=\phi$ and $\delta\theta_{i}=\delta\theta,$
so the holonomy is:
\begin{eqnarray*}
^{3}H & = & I\cos^{3}\frac{\delta\theta}{2}+\left(\cos^{2}\frac{\delta\theta}{2}\,\sin\frac{\delta\theta}{2}\right)\left(\sum_{i=1}^{3}\vec{J}_{i}\right)\cdot\hat{\bar{\sigma}}+\left(\cos\frac{\delta\theta}{2}\,\sin^{2}\frac{\delta\theta}{2}\right)\sum_{i,j=1,i\neq j}^{3}\left(\vec{J}_{i}\cdot\hat{\bar{\sigma}}\right)\left(\vec{J}_{j}\cdot\hat{\bar{\sigma}}\right)+\sin^{3}\frac{\delta\theta}{2}\prod_{i=1}^{3}\left(\vec{J}_{i}\cdot\hat{\bar{\sigma}}\right),
\end{eqnarray*}
Therefore, the 3D intrinsic curvature for isosceles tetrahedra case
is: 
\[
^{3}R=\delta\Theta=2\arccos\left(\cos^{3}\frac{\delta\theta}{2}-3\cos\phi\,\cos\frac{\delta\theta}{2}\,\sin^{2}\frac{\delta\theta}{2}+\left(\textrm{vol}\right)\sin^{3}\frac{\delta\theta}{2}\right).
\]

\subsubsection{2D extrinsic curvature}

Equation (\ref{eq:2.12}) is a linear relation between the trace of
3D holonomy with the 2D holonomy and its 'residual' part $T.$ Inserting
(\ref{eq:bb}) and (\ref{eq:cc}) to (\ref{eq:2.12}), we could obtain
$\textrm{tr}T,$ but this is\textit{ not} the extrinsic curvature, since
$T=S+\left[K,K\right]+\left(I_{3}-I_{2}\right).$ Next, we would like
to obtain an angle describing the extrinsic curvature $K$. We do\textit{
not} expect the relation between this angle with $^{3}R$ and $^{2}R$
to be linear, since linearity is held on the holonomy representation
of Gauss-Codazzi (\ref{eq:2.10}), while the relation between their
rotation angle $^{3}R$ and $^{2}R$ does\textit{ not} need to be
linear.

Consider the isosceles tetrahedron case. On each segment
of the tetrahedron, lies the 3D deficit angle $\delta\theta.$ This
deficit angle is defined as: 
\[
\delta\theta=2\pi-\left(\theta+\alpha\right),
\]
where $\theta$ is the \textit{internal} dihedral angle, and $\alpha$
is the \textit{external} dihedral angle coming from the dihedral angles
of \textit{other} tetrahedra (remember that the full discretized $(2+1)$
picture is the 1-4 Pachner moves). Let's introduce the quantity 
\begin{equation}
\theta^{'}=\alpha-\pi.\label{eq:key}
\end{equation}
For the case where $M_{\Delta}$ is flat, $\delta\theta=0$. This
causes $\theta+\alpha=2\pi,$ and using definition (\ref{eq:key}),
we obtain: 
\[
\theta+\theta'=\pi.
\]
In this flat case, it is clear that $\theta'$ is the angle between
the normals of two triangles, see FIG. 5.

\begin{figure}
\centerline{\includegraphics[height=3cm]{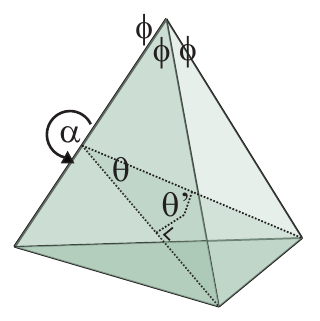}} \caption{From the dihedral angle relation, we could obtain the \textit{internal}
dihedral angle $\theta$ on a segment. $\alpha$ is the \textit{external}
dihedral angle, while $\theta'$ is the angle between \textit{normals}
of the two triangles.}
\end{figure}

Therefore, $\theta'$ is in accordance with the definition of extrinsic
curvature in (\ref{eq:2.3}), where $K$ is defined as the covariant
derivative of the normal $n$ to the hypersurface $\Sigma$. Because
of this reason, we define $\theta^{'}$ as the 2D \textit{extrinsic
curvature}, since in a general curved case, it will inherit the curvature
of the 3D manifold.

\subsection{Discrete Gauss-Codazzi equation in (2+1) dimension}

A simpler but equivalent way to obtain the 3D instrinsic curvature
is to use the purely geometrical picture of the discretized manifold.
See FIG. 6.

\begin{figure}
\centerline{\includegraphics[height=6cm]{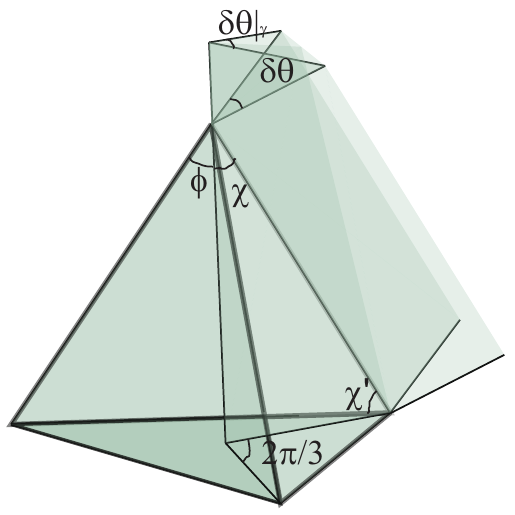}} \caption{The 3D deficit angle on each segment can be thought as a 'piece of
space' inserted along the hinge. The 'size' of this 'piece of space'
is described by the deficit angle $\delta\theta,$ which is located
on the plane normal to the hinge. The total deficit angle $\delta\Theta$
obtained by adding three rotation in (\ref{eq:hol}), is located in
the plane normal to an 'artificial' hinge in the direction $\hat{k}$,
since we use isosceles tetrahedron. Therefore, projecting $\delta\theta$
to this plane, we obtain $\left.\delta\theta\right|_{\gamma}$.}
\end{figure}

From Figure 6, we obtain the projected 3D intrinsic curvature of each
segment on the 'artificial' segment dual to the total plane of rotation
as: 
\begin{equation}
\cos^{2}\frac{\left.\delta\theta\right|_{\gamma}}{2}=\frac{\sin^{2}\chi'\cos^{2}\frac{\delta\theta}{2}}{1-\cos^{2}\chi'\cos^{2}\frac{\delta\theta}{2}}.\label{eq:2.17}
\end{equation}
Using 
\[
\vec{J}_{i}=\sin\chi'\hat{k}+\cos\chi\hat{r}_{i},
\]
we obtain: 
\[
\left\langle \vec{J}_{i},\vec{J}_{j}\right\rangle =\cos\phi=\sin^{2}\chi'+\cos^{2}\chi'\cos^{2}\frac{2\pi}{3},
\]
or: 
\begin{eqnarray}
\cos^{2}\chi' & = & \frac{2}{3}\left(1-\cos\phi\right),\label{eq:2.17a}\\
\sin^{2}\chi' & = & \frac{1}{3}\left(1+2\cos\phi\right),\label{eq:2.17b}
\end{eqnarray}
since we use isosceles tetrahedron. All the curvatures: 3D intrinsic,
2D intrinsic, and 2D extrinsic curvatures are contained in (\ref{eq:2.17}),
through: 
\begin{equation}
\left.\delta\theta\right|_{\gamma}=\frac{\delta\Theta}{3}=\frac{^{3}R}{3},\label{eq:2.17c}
\end{equation}
\begin{equation}
\phi=\frac{2\pi-\delta\phi}{3}=\frac{2\pi-{}^{2}R}{3}\label{eq:2.17d}
\end{equation}
\begin{equation}
\delta\theta=\pi-\left(\theta+\theta'\right)=\pi-\left(\arccos\left(\frac{\cos\phi}{1+\cos\phi}\right)+K\right).\label{eq:2.17e}
\end{equation}
Using the half-angle formula and inserting (\ref{eq:2.17a})-(\ref{eq:2.17e})
to (\ref{eq:2.17}), we obtain the relation between angles of the
discrete Gauss-Codazzi equation for a special discretization using
isosceles tetrahedra: 
\begin{equation}
1-\cos\left(K+\arccos\left(\frac{\cos\frac{2\pi-{}^{2}R}{3}}{1+\cos\frac{2\pi-{}^{2}R}{3}}\right)\right)=\frac{1+\cos\frac{^{3}R}{3}}{1-\frac{1}{3}\left(1-\cos\frac{^{3}R}{3}\right)\left(1-\cos\frac{2\pi-{}^{2}R}{3}\right)}.\label{eq:GC}
\end{equation}
Together with their planes of rotation: $j$ in (\ref{eq:2.11a})
for $^{2}R$ and $\hat{k}\cdot\bar{\sigma}$ for $^{3}R$ (since we
use isosceles tetrahedron), (\ref{eq:GC}) is the Gauss-Codazzi equation
in plane-angle representation, which contains same amount of information
provided by the linear Gauss-Codazzi equation in holonomy representation
in (\ref{eq:2.10}).

\section{Spin network states and 3D loop quantum gravity} \label{IV}

\subsection{2-complex and its slice}

In this section, we apply the results obtained above to loop quantum gravity (LQG). 
We will briefly review the quantization of gravity
in first order formalism using the loop representation. From the real
space representation, we move to the (Hodge) dual space \cite{key-8}.
(The dual space is introduced in analogy to what is usually done in 
higher dimension.) In LQG, the fundamental geometrical object is
the \textit{2-complex}. A 2-complex is a 2-dimensional geometrical
object dual to the discrete manifold
$M_{\triangle}.$ Let us call the 2-complex as $\star M_{\triangle}$.
We can take a slice on a 2-complex, which is a \textit{1-complex}
called as $\star\Sigma_{\triangle}$, dual to the discrete hypersurface
$\Sigma_{\triangle}$ of $M_{\triangle}.$ A 1-complex is a \textit{graph}
consisting oriented lines called \textit{link}s $l$ and points
called \textit{nodes} $n$. The variables attached to the graphs
are the variables coming from first order formalism: the holonomy
$U_{l}\in SU(2)$ and algebra $J_{l}\in\mathfrak{su(2)}^{*}$ on each
links $l$ of the graph. A graph with elements of $SU(2)\times\mathfrak{su(2)}^{*}$
is called a \textit{spin network}. In the canonical quantization,
we promote $\left\{ U_{l},J_{l}\right\} $ to operators $\left\{ \hat{U}_{l},\hat{J}_{l}\right\} $.
The space $SU(2)\times\mathfrak{su(2)}^{*}$ is the phase-space of
LQG, and the pair $\left\{ \hat{U}_{l},\hat{J}_{l}\right\} $ are
the pair conjugate to each other, satisfying a quantized-algebra structure
as follow \cite{key-3,key-8,key-11}: 
\begin{eqnarray*}
\left[\hat{U}_{l},\hat{U}_{l'}\right] & = & 0\\
\left[\hat{U}_{l},\hat{J}_{l'}^{i}\right] & = & -\delta_{ll'}\bar{\sigma}^{i}\hat{U}_{l}\\
\left[\hat{J}_{l}^{i},\hat{J}_{l'}^{j}\right] & = & -\delta_{ll'}\varepsilon_{\; k}^{ij}\hat{J}_{l}^{k}.
\end{eqnarray*}
On each node, the gauge invariant condition must be satisfied: 
\[
\hat{\mathcal{G}}\left|\psi\right\rangle =\sum_{l\in n}\hat{J}_{l}\left|\psi\right\rangle =0.
\]
This condition selects an invariant Hilbert space $\mathcal{K}=L_{2}\left[SU(2)^{l}/SU(2)^{n}\right]\ni\left|\psi\right\rangle $
of a closed triangle formed by $\left\{ \hat{J}_{1},\hat{J}_{2},\hat{J}_{3}\right\} $
\cite{key-3,key-8,key-11}. As an example for our case, see FIG. 7.

\begin{figure}
\centerline{\includegraphics[height=6cm]{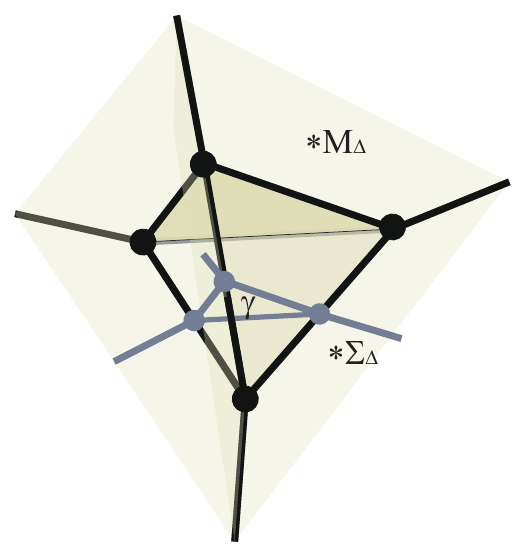}} \caption{The figure above is a 2-complex dual to the geometric figure in FIG.
1. Suppose we have a 3-dimensional curved manifold $M_{\Delta}$ discretized
by four tetrahedron as the first figure above (in flat case it is
known as 1-4 Pachner move). Then we take an embedded slice $\Sigma_{\triangle}$
as the surface of one tetrahedron (the dark blue surface discretized
by three triangle). Going to the dual space picture, $M_{\Delta}$
is dual to a 2-complex $\star M_{\Delta}$ called as the 3D \textit{bubble}
\cite{key-8} (its geometry is defined by black lines and black points),
while $\Sigma_{\triangle}$ is dual to a 1-complex $\star\Sigma_{\triangle}$
called as 2D \textit{bubble} (defined by purple lines and point).
Clearly, the 2D bubble $\star\Sigma_{\triangle}$ is embedded on the
3D bubble $\star M_{\Delta}$. The bubble graph define complete curvature
of the portion of space.}
\end{figure}

\subsection{Curvature operators}

Given a 2D bubble graph with a phase space variable
on each link, we can construct the operators of 3D, 2D intrinsics
and 2D extrinsic curvature from these variables.

\begin{figure}
\centerline{\includegraphics[height=2.5cm]{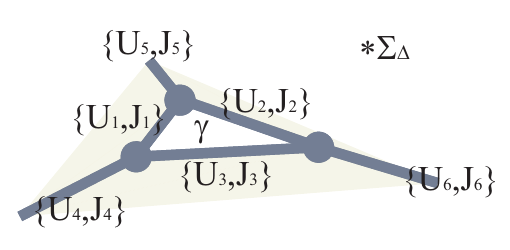}} \caption{Taking only the slice on the dual space representation, we obtain
the spin network $\star\Sigma_{\triangle}.$ $\gamma$ is the loop
chosen. Inside $\gamma$ is where the point $p$ located in the real
space representation.}
\end{figure}

We construct these operators using the basic phase-space operators,
for example, the angle operator from the phase-space variables \cite{key-12}: 
\[
\hat{\phi}_{i}=\arccos\left(\frac{\left|\hat{J}_{j}\right|^{2}+\left|\hat{J}_{k}\right|^{2}-\left|\hat{J}_{i}\right|^{2}}{2\left|\hat{J}_{j}\right|\left|\hat{J}_{k}\right|}\right),\quad i,j,k\in n.
\]
From this angle operator, we obtain the 2D intrinsic curvature
along loop $\gamma$ as: 
\[
^{2}\hat{R}=2\pi-\sum_{i\subseteq\gamma}\hat{\phi}_{i},
\]
which is clearly the deficit angle on a point in the real space representation.
See FIG. 8.

The 3D intrinsic curvature can be obtained from the holonomy around
a loop. Remember that the direct geometrical interpretation of spacetime
can only be obtained from second order formalism, while in LQG, the
fundamental variables \textit{comes from first order formalism}: the
holonomy $U$ comes from the curvature 2-form $F$, instead of the
Riemannian curvature $R$ (see Section 1). The relation between $F$
and $R$, subjected to the torsionless condition is: 
\begin{equation}
R=e\left(F\right),\label{eq:send}
\end{equation}
with $e$ is the local trivialization map between them.
Contracting (\ref{eq:send}) with an infinitesimal area inside the
loop (carried by the derivation in Section 1) gives: 
\[
H=e\left(U\right),\quad U=e^{-1}\left(H\right),
\]
since $e$ is a diffeomorphism and must have inverse.

The 3D intrinsic curvature is $^{3}R\sim\textrm{tr}H$, with $H$
is the holonomy coming from the contraction of Riemanian curvature
$R$ with an infinitesimal area. But since $e$ is a diffeomorphism
and trace of the holonomy is \textit{invariant} under diffeomorphism
and gauge transformation, we obtain $\textrm{tr}H=\textrm{tr}U.$
Therefore, we can write the 3D intrinsic curvature operator as:

\begin{equation}
^{3}\hat{R}=2\arccos\frac{\textrm{tr}\hat{U}_{1}\hat{U}_{2}\hat{U}_{3}}{2}.\label{eq:r3}
\end{equation}
The extrinsic curvature operator can be obtained directly from the
discrete Gauss-Codazzi equation (\ref{eq:GC}):

\[
\hat{K}=\arccos\left(1-\frac{1+\cos\frac{^{3}\hat{R}}{3}}{1-\frac{1}{3}\left(1-\cos\frac{^{3}\hat{R}}{3}\right)\left(1-\cos\frac{2\pi-{}^{2}\hat{R}}{3}\right)}\right)-\arccos\left(\frac{\cos\frac{2\pi-{}^{2}\hat{R}}{3}}{1+\cos\frac{2\pi-{}^{2}\hat{R}}{3}}\right).
\]

\subsection{The semi-classical limit}

In this section, we show that the linearity of the relation between $\textrm{tr}\,^{3}H$
and $\textrm{tr}\,^{2}H$ in equation (\ref{eq:2.12}) is recovered 
from the spin network calculation in the semi-classical limit, by
dropping the Maslov phase. Let us take a simple spin network illustrated
in FIG. 9. 
\begin{figure}
\centerline{\includegraphics[height=3.5cm]{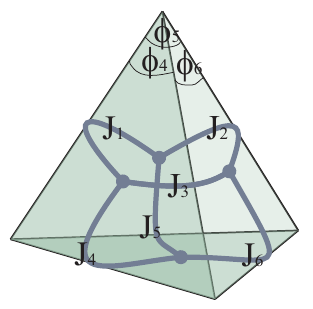}} \caption{A simple spin network $\star\Sigma_{\Delta}$ (the graph in purple), which
will give a geometrical interpretation of a 'fuzzy' quantum tetrahedron in the semi-classical limit. This spin network can be obtained from
the spin network in FIG. 8, by identifying their open links together. In 3D theory, the nodes (purple points) are dual to triangles, and
the links (purple lines) are dual to segments.}
\end{figure}
 For simplicity in our calculation, we use a closed graph as an example, but the result is valid for any spin network graph. The vector and
covector state of the spin network $\star\Sigma_{\Delta}$ in the algebra
representation are:
\begin{equation}
\left|\psi\right\rangle =\sum_{\binom{m_{1}\ldots m_{6}}{n_{1}..n_{6}}}i^{m_{6}m_{3}n_{2}}i^{m_{5}m_{2}n_{1}}i^{m_{4}m_{1}n_{3}}i^{n_{6}n_{5}n_{4}}\left|j_{1}m_{1}n_{1}\ldots j_{6}m_{6}n_{6}\right\rangle ,\label{eq:vect}
\end{equation}
\begin{equation}
\left\langle \psi\right|=\sum_{\binom{m_{1'}\ldots m_{6'}}{n_{1'}..n_{6'}}}i^{m_{6}'m_{3}'n_{2}'}i^{m_{5}'m_{2}'n_{1}'}i^{m_{4}'m_{1}'n_{3}'}i^{n_{6}'n_{5}'n_{4}'}\left\langle j_{1}'m_{1}'n_{1}'\ldots j_{6}'m_{6}'n_{6}'\right|,\label{eq:covect}
\end{equation}
with $i$ is the intertwinner attached on the node. See \cite{key-8,key-11,key-13}
for a detail explanation about the spin network state. Changing the representation basis using:
\[
\left.\left\langle U\right|jmn\right\rangle =D_{mn}^{j}\left(U\right),
\]
with $D_{mn}^{j}\left(U\right)$ is the matrix representation of $U\in SU(2)$
in $(2j+1)$-dimension, we can write the vector and covector states in the group representation basis:
\[
\left|\psi\right\rangle =\sum_{\binom{m_{1}\ldots m_{6}}{n_{1}..n_{6}}}i^{m_{6}m_{3}n_{2}}i^{m_{5}m_{2}n_{1}}i^{m_{4}m_{1}n_{3}}i^{n_{6}n_{5}n_{4}}\intop dU_{1}..dU_{6}\prod_{i=1}^{6}D_{m_{i}n_{i}}^{j_{i}}\left(U_{i}\right)\left|U_{1}..U_{6}\right\rangle ,
\]
\[
\left\langle \psi\right|=\sum_{\binom{m_{1'}\ldots m_{6'}}{n_{4'}..n_{6'}}}i^{m_{6}'m_{3}'n_{2}'}i^{m_{5}'m_{2}'n_{1}'}i^{m_{4}'m_{1}'n_{3}'}i^{n_{6}'n_{5}'n_{4}'}\intop dU_{1}'..dU_{6}'\prod_{i=1}^{6}D_{m_{i}'n_{i}'}^{j_{i}'}\left(U_{i}'\right)\left\langle U_{1}'..U_{6}'\right|.
\]

We choose a specific closed path on spin network $\star\Sigma_{\Delta}$, say, loop $\gamma$, defined by links $\left\{ 1-2-3\right\} $ (See FIG. 9). The holonomy $U=U_{1}U_{2}U_{3}$ are attached along $\gamma$, therefore we have a well-defined operator related to the 3D intrinsic
curvature operator (\ref{eq:r3}) on the loop: 
\begin{equation}
\textrm{tr}\,^{3}H=\underset{2\cos\frac{^{3}\hat{R}}{2}}{\underbrace{\textrm{tr}U}}=\textrm{tr}\left(U_{1}U_{2}U_{3}\right)=\sum_{m,n}\delta_{mn}D_{mn}^{j}\left(U_{1}U_{2}U_{3}\right).\label{eq:x}
\end{equation}
Since the matrix representation satisfies:
\begin{equation}
\left\langle j,m\right|\hat{D}\left(U_{1}U_{2}U_{3}\right)\left|j,n\right\rangle =D_{mn}^{j}\left(U_{1}U_{2}U_{3}\right)=\sum_{k,l}D_{mk}^{j}\left(U_{1}\right)D_{kl}^{j}\left(U_{2}\right)D_{ln}^{j}\left(U_{3}\right),\label{eq:y}
\end{equation}
we can write
\begin{equation}
\textrm{tr}\,^{3}H=\sum_{k,l,m,n}\delta_{mn}D_{mk}^{j}\left(U_{1}\right)D_{kl}^{j}\left(U_{2}\right)D_{ln}^{j}\left(U_{3}\right).\label{eq:z}
\end{equation}
The geometrical interpretation of the spin-$j$ in (\ref{eq:x})-(\ref{eq:z})
is the measure on the 'artificial' hinge dual to the plane of rotation
inside loop $\gamma.$ See FIG. 6 and Section III C.

Acting $\textrm{tr}\,^{3}H$ with the spin network states gives:
\begin{align*}
\left\langle \psi\right|\textrm{tr}\,^{3}H\left|\psi\right\rangle = & \left(\sum_{m_{6},l,m_{3},n_{2},m_{3'},n_{2'}}i^{m_{6}m_{3}n_{2}}i^{m_{6}m_{3}'n_{2}'}i^{m_{3}lm_{3}'}i^{n_{2}ln_{2}'}\right)\left(\sum_{m_{5},k,m_{2},n_{1},m_{2'},n_{5'}}i^{m_{5}m_{2}n_{1}}i^{m_{5}m_{2}'n_{1}'}i^{m_{2}km_{2}'}i^{n_{1}kn_{1}'}\right)\times..\\
 & \;...\times\left(\sum_{m_{4},m,m_{1},n_{3},m_{1'},n_{3'}}i^{m_{4}m_{1}n_{3}}i^{m_{4}m_{1}'n_{3}'}i^{n_{3}mn_{3}'}i^{m_{1}mm_{1}'}\right)\left(\sum_{n_{4}..n_{6}}i^{n_{6}n_{5}n_{4}}i^{n_{6}n_{5}n_{4}}\right),
\end{align*}
but the first three-terms in the parantheses are only the Wigner 6j-symbols:
\begin{equation}
\left\langle \psi\right|\textrm{tr}\,^{3}H\left|\psi\right\rangle =\underset{\prod_{i=1}^{3}\left\{ 6j\right\} _{i}}{\underbrace{\left\{ \begin{array}{ccc}
j_{6} & j_{3} & j_{2}\\
j_{3}' & j_{2}' & j
\end{array}\right\} \left\{ \begin{array}{ccc}
j_{5} & j_{2} & j_{1}\\
j_{2}' & j_{1}' & j
\end{array}\right\} \left\{ \begin{array}{ccc}
j_{4} & j_{1} & j_{3}\\
j_{1}' & j_{3}' & j
\end{array}\right\} }}\left(\sum_{n_{4}..n_{6}}i^{n_{6}n_{5}n_{4}}i^{n_{6}n_{5}n_{4}}\right).\label{eq:huhu}
\end{equation}

We are ready to take the semi-classical limit by setting all the
$j_{i}$'s to be large. Using the result proved by Roberts \cite{key-14,key-15},
for large spin $j,$ the 6j-symbol can be approximated as follow:
\[
\left\{ 6j\right\} =\left\{ \begin{array}{ccc}
j_{1} & j_{2} & j_{3}\\
j_{4} & j_{5} & j_{6}
\end{array}\right\} \approx\frac{1}{\sqrt{12\pi\left|V\right|}}\cos\left(\sum_{i=1}^{6}j_{i}\theta_{i}-\frac{\pi}{4}\right),
\]
where the six $j_{i}$ constructs a tetrahedron, $\theta_{i}$ are
the internal dihedral angle on segment $j_{i},$ and $\left|V\right|$
is the volume of the tetrahedron. The $\frac{\pi}{4}$ term is known
as the Maslov phase. Using this result to our case, we obtain:
\begin{align}
\left\{ \begin{array}{ccc}
j_{6} & j_{3} & j_{2}\\
j_{3}' & j_{2}' & j
\end{array}\right\}  & \approx\frac{1}{\sqrt{12\pi\left|V\right|}}\cos\left(j_{6}\theta_{6}+j_{3}\theta_{3}+j_{2}\theta_{2}+j_{3}'\theta_{3}'+j_{2}'\theta_{2}'+j\theta_{l}-\frac{\pi}{4}\right)\nonumber \\
 & \approx\frac{1}{\sqrt{12\pi\left|V\right|}}\cos\left(j_{6}\theta_{6}+2j_{3}\theta_{3}+2j_{2}\theta_{2}+j\theta_{l}-\frac{\pi}{4}\right).\label{eq:hu}
\end{align}
where the last step we use the fact that $j_{i}=j_{i}',$ noting that
$j_{i}'$ comes from the covector state (\ref{eq:covect}). This fact also
cause $\theta_{i}=\theta_{i}'$. See FIG. 10. 
\begin{figure}
\centerline{\includegraphics[height=3.5cm]{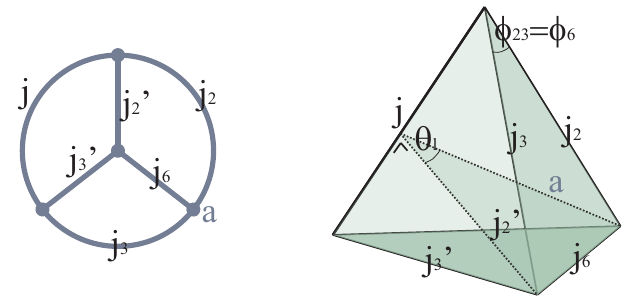}} \caption{The 6j-symbol $\left\{ \protect\begin{array}{ccc}
j_{6} & j_{3} & j_{2}\protect\\
j_{3}' & j_{2}' & j
\protect\end{array}\right\} $ can be represented using a graph (left), where it can be translated
into its dual (right) which is a tetrahedron. Node $a$ on the graph
is the triangle $a$ on the tetrahedron. $\theta_{l}$ is the dihedral
angle on $j.$ }
\end{figure}
 Remember that $j$ is the measure on the 'artificial' hinge
inside loop $\gamma$, in other words, the spin number $j$ is the
'length' of the artificial hinge. Along this artificial hinge, there
is a point $p$ where the 2D deficit angle is located. Taking $j_{i}\ggg j,$
we have: 

\[
\theta_{6}\approx0,\:\theta_{3}\approx\frac{\pi}{2},\:\theta_{2}\approx\frac{\pi}{2},\:\theta_{l}\approx\phi_{23}=\phi_{6},
\]
See FIG. 10. Using this approximation, (\ref{eq:hu}) becomes:
\[
\left\{ \begin{array}{ccc}
j_{6} & j_{3} & j_{2}\\
j_{3}' & j_{2}' & j
\end{array}\right\} \approx\frac{1}{\sqrt{12\pi\left|V_{1}\left(j_{6},j_{3},j_{2},j\right)\right|}}\cos\left(\left(j_{3}+j_{2}\right)\pi+j\phi_{6}-\frac{\pi}{4}\right).
\]
Doing the same way for the other two 6j-symbols, we obtain:
\[
\left\{ \begin{array}{ccc}
j_{5} & j_{2} & j_{1}\\
j_{2}' & j_{1}' & j
\end{array}\right\} \approx\frac{1}{\sqrt{12\pi\left|V_{2}\left(j_{5},j_{2},j_{1},j\right)\right|}}\cos\left(\left(j_{2}+j_{1}\right)\pi+j\phi_{5}-\frac{\pi}{4}\right),
\]
\[
\left\{ \begin{array}{ccc}
j_{4} & j_{1} & j_{3}\\
j_{1}' & j_{3}' & j
\end{array}\right\} \approx\frac{1}{\sqrt{12\pi\left|V_{3}\left(j_{4},j_{1},j_{3},j\right)\right|}}\cos\left(\left(j_{1}+j_{3}\right)\pi+j\phi_{4}-\frac{\pi}{4}\right).
\]
By writing the cosine function using the Euler formula:
\begin{eqnarray*}
\cos x & = & \frac{1}{2}\left(\exp ix+\exp-ix\right),
\end{eqnarray*}
we can write the products of the three 6j-symbols as:
\begin{eqnarray}
\prod_{i=1}^{3}\left\{ 6j\right\} _{i} & \approx & \frac{1}{4}\frac{1}{12\pi\sqrt{12\pi\left|V_{1}\right|\left|V_{2}\right|\left|V_{3}\right|}}\times\left(\:\cos\left(2\left(j_{1}+j_{2}+j_{3}\right)\pi+j\left(\phi_{4}+\phi_{5}+\phi_{6}\right)-\frac{3\pi}{4}\right)+...\right.\nonumber \\
 &  & \qquad...+\cos\left(2j_{3}\pi+j\left(\phi_{4}-\phi_{5}+\phi_{6}\right)-\frac{\pi}{4}\right)+\cos\left(2j_{2}\pi+j\left(-\phi_{4}+\phi_{5}+\phi_{6}\right)-\frac{\pi}{4}\right)+...\nonumber \\
 &  & \qquad...+\left.\cos\left(2j_{1}\pi+j\left(\phi_{4}+\phi_{5}-\phi_{6}\right)-\frac{\pi}{4}\right)\,\right).\label{eq:huh}
\end{eqnarray}
From this point, we take the sum of $j_{1}+j_{2}+j_{3}$ to be an integer/natural
number (we discard the half-integer possibility) and choose the spin number $j=1,$ then we could write:
\begin{align}
\prod_{i=1}^{3}\left\{ 6j\right\} _{i} & \approx\frac{1}{4}\frac{1}{12\pi\sqrt{12\pi\left|V_{1}\right|\left|V_{2}\right|\left|V_{3}\right|}}\:\cos\left(\delta\phi-\frac{3\pi}{4}\right)+\mathcal{O},\label{eq:huhuhu}
\end{align}
where we denoted the remaining terms by $\mathcal{O}$ and using $\phi_{4}+\phi_{5}+\phi_{6}=2\pi-\delta\phi$
(see FIG. 9). Inserting (\ref{eq:huhuhu}) to (\ref{eq:huhu}), we
obtain:
\begin{align*}
\left\langle \cos\frac{^{3}\hat{R}}{2}\right\rangle & \sim\left\langle \cos\left(^{2}\hat{R}-\frac{3\pi}{4}\right)\right\rangle.
\end{align*}
The linearity relation (\ref{eq:2.12}) can be recovered by dropping the Maslov phase:
\begin{align*}
\left\langle \textrm{tr}\,^{3}H\right\rangle  & \sim\left\langle \textrm{tr}\,^{2}H\right\rangle .
\end{align*}

An important point is we have a freedom to choose the spin number
$j,$ since this spin number comes from (\ref{eq:x}), which is the
dimension of the $SU(2)$ representation. If we set $j=0,$ then,
the volume of the 6j tetrahedra $\left|V_{i}\right|$ will be zero
and relation (\ref{eq:huh}) will diverge, given any $j_{i}'$s. But
since the range of spin $j$ is $j\geq\frac{1}{2}$, with $j$ multiple
of half, then the spin number $j$ acts as an ultraviolet cut-off which
prevents relation (\ref{eq:huh}) from divergence. 

\section{Conclusion} \label{V}

We have obtained the Gauss-Codazzi equation for a discrete (2+1)-dimensional manifold, discretized by isosceles tetrahedra. We have studied the definition of extrinsic curvature in the discretized context. With definitions of the 3-dimensional intrinsic, 2-dimensional intrinsic, and 2-dimensional extrinsic curvature in the discrete picture, we have promoted them to curvature operators, acting on the spin network states of (2+1)-dimensional loop quantum gravity. In the semi classical limit, we have shown that the linearity between $\textrm{tr}\,^{3}H$ and $\textrm{tr}\,^{2}H$ in equation (\ref{eq:2.12}) is recovered in the spin network calculation, by dropping the Maslov phase. There exist a natural ultraviolet cut-off which prevents the discrete Gauss-Codazzi equation from divergences in the semi-classical limit. 

\bibliographystyle{apsrev4-1}

   \bibliography{library}
\begin{thebibliography}{10}

\bibitem{key-1}S. M. Carroll. \textit{Spacetime and geometry}: \textit{An introduction to general relativity}. San Francisco,
CA, USA. Addison-Wesley. ISBN 0-8053-8732-3. 2004.

\bibitem{key-2}T. Regge, R. M. Williams. \textit{Discrete structures in gravity}. J. Math. Phys. \textbf{41}, 3964 (2000). \href{ http://arxiv.org/pdf/gr-qc/0012035v1.pdf}{arXiv:gr-qc/0012035v1}.

\bibitem{key-3}P. Don$\acute{\textrm{a}}$, S. Speziale.
\textit{Introductory lectures to loop quantum gravity}. (2010). \href{http://arxiv.org/abs/1007.0402}{arXiv:gr-qc/1007.0402}.

\bibitem{key-4}J. W. Barrett and I. Naish-Guzman.
\textit{The Ponzano-Regge model}. Class. Quant. Grav. \textbf{26}:
155014 (2009). \href{http://arxiv.org/abs/0803.3319}{arXiv:gr-qc/0803.3319}.

\bibitem{key-5}T. Regge. \textit{General relativity
without coordinates}. Nuovo Cim. \textbf{19} (1961) 558. \href{ http://www.signalscience.net/files/Regge.pdf}{ http://www.signalscience.net/files/Regge.pdf}.

\bibitem{key-6}R. Arnowitt, S. Deser, C. Misner.
\textit{Dynamical Structure and Definition of Energy in General Relativity}.
Phys. Rev. \textbf{116} (5): 1322-1330. (1959).

\bibitem{key-8}C. Rovelli, F. Vidotto. \textit{Covariant
Loop Quantum Gravity: An Elementary Introduction to Quantum Gravity
and Spinfoam Theory}. UK. Cambridge University Press. ISBN 978-1-107-06962-6.
2015.

\bibitem{key-7}J. C. Baez, J. P. Muniain. \textit{Gauge
Fields, Knots, and Gravity. }Series on Knots and Everything: vol \textbf{4}.
World Scientific Pub Co Ltd. ISBN 9789810220341. 1994.

\bibitem{key-9}H. W. Hamber. \textit{Quantum Gravitation:
The Feynman Path Integral Approach}. Springer-Verlag. ISBN 978-3-540-85292-6.
2009.

\bibitem{key-10}B. C. Hall. \textit{An Elementary
Introduction to Groups and Representations}. Graduate Texts in Mathematics
\textbf{222} (2003). Springer-Verlag. \href{http://arxiv.org/pdf/math-ph/0005032v1.pdf}{arXiv:math-ph/0005032v1}.

\bibitem{key-11}C. Rovelli. \textit{Zakopane lectures
on loop gravity}. (2011). \href{http://arxiv.org/abs/1102.3660}{arXiv:gr-qc/1102.3660}.

\bibitem{key-12}M. Seifert. \textit{Angle and Volume
Studies in Quantized Space}. \href{ http://arxiv.org/abs/gr-qc/0108047}{arXiv:gr-qc/0108047}.

\bibitem{key-13}S. A. Major. A \textit{Spin Network
Primer}. Am. J. Phys. \textbf{67} (1999) 972-980. \href{ http://arxiv.org/abs/gr-qc/9905020}{arXiv:gr-qc/9905020}.

\bibitem{key-14}J. Roberts. \textit{Classical 6j-symbols
and the tetrahedron.} Geom. Topol. \textbf{3} (1999). \href{http://arxiv.org/abs/math-ph/9812013}{arXiv:math-ph/9812013}.

\bibitem{key-15}G. Ponzano and T. Regge. \textit{Semiclassical
Limit of Racah Coefficients}. Spectroscopy and Group Theoretical Methods
in Physics.  Amsterdam. pp. 1-58 (1968).

\end{thebibliography}
\end{document}